\documentclass{article}
\usepackage{spconf,amsmath,graphicx}

\usepackage[colorlinks=true,allcolors=black]{hyperref}
\usepackage{url}
\usepackage{comment}

\newcommand{\bhline}[1]{\noalign{\hrule height #1}}

\usepackage{multirow}

\usepackage{amsmath,amsfonts,bm}
\usepackage{amsthm}

\def\eqref#1{equation~\ref{#1}}
\def\beqref#1{(\ref{#1})}
\def\1{\bm{1}}

\DeclareMathAlphabet{\mathsfit}{\encodingdefault}{\sfdefault}{m}{sl}
\SetMathAlphabet{\mathsfit}{bold}{\encodingdefault}{\sfdefault}{bx}{n}

\newcommand{\E}{\mathbb{E}}

\newcommand{\Js}{\mathcal{J}}

\usepackage{color,xcolor}
\definecolor{c_yuhta}{rgb}{0.831,0.184,0.494}

\definecolor{c_takashi}{rgb}{0.184,0.494,0.831}

\title{BigVSAN: Enhancing GAN-based Neural Vocoders with \\ Slicing Adversarial Network}
\name{
Takashi Shibuya${}^1$ \qquad
Yuhta Takida${}^1$ \qquad
Yuki Mitsufuji${}^{1,2}$
}
\address{
${}^1$Sony AI, Tokyo, Japan,\quad
${}^2$Sony Group Corporation, Tokyo, Japan
}

\begin{document}
\ninept
\maketitle
\begin{abstract}
Generative adversarial network (GAN)-based vocoders have been intensively studied because they can synthesize high-fidelity audio waveforms faster than real-time. However, it has been reported that most GANs fail to obtain the optimal projection for discriminating between real and fake data in the feature space. In the literature, it has been demonstrated that slicing adversarial network (SAN), an improved GAN training framework that can find the optimal projection, is effective in the image generation task. In this paper, we investigate the effectiveness of SAN in the vocoding task. For this purpose, we propose a scheme to modify least-squares GAN, which most GAN-based vocoders adopt, so that their loss functions satisfy the requirements of SAN. Through our experiments, we demonstrate that SAN can improve the performance of GAN-based vocoders, including BigVGAN, with small modifications. Our code is available at \url{https://github.com/sony/bigvsan}.

\end{abstract}
\section{Introduction}
\label{sec:intro}

Speech synthesis technology has made rapid progress with the development of neural networks~\cite{shen2018natural,ren2021fastspeech}. Most speech synthesis systems adopt a two-stage pipeline that 1) predicts the intermediate representation (e.g., mel-spectrograms) from text and then 2) transforms the intermediate representation to audio. In this work, we focus on a latter-stage model called a vocoder, which synthesizes waveforms from mel-spectrograms.

Various approaches have been studied to improve the speech synthesis quality of vocoders. Deep generative models have demonstrated a particularly high performance in synthesizing waveforms, including autoregressive models~\cite{oord2016wavenet,kalchbrenner2018efficient}, generative adversarial network (GAN)-based models~\cite{kumar2019melgan,yamamoto2020parallel,kong2020hifigan}, flow-based models~\cite{oord2018parallel,ping2018clarinet,prenger2018waveglow,kim2019flowavenet,ping2020waveflow,lee2020nanoflow}, and diffusion-based models~\cite{chen2021wavegrad,kong2021diffwave,lee2022priorgrad,koizumi2022specgrad,takahashi2023hierarchical}.
Although deep generative models struggle with simultaneously addressing the three key requirements of high sample quality, diversity (mode coverage), and fast sampling~\cite{xiao2022tackling}, diversity is not as crucial for vocoders as for unconditional image generation models because vocoders are required to synthesize a waveform corresponding to a given mel-spectrogram. 
Thus, since GAN can generate high-quality samples rapidly, it is a powerful and reasonable tool that vocoders can rely on. BigVGAN~\cite{lee2023bigvgan} has recently demonstrated the capability of generating high-fidelity waveforms conditioned on mel-spectrograms much faster than real-time on a single GPU, achieving state-of-the-art performance.

The discrimination process in GANs essentially consists of the following two steps~\cite{takida2023san}: 1) extracting features by neural networks and 2) projecting the features with linear operations into one-dimensional space to judge if each input is real or fake with a scalar. Most studies on GAN-based vocoders have enhanced the feature extraction part by proposing new discriminator architectures (see Sec.~\ref{sec:relateed_work}), but with the exception of Wasserstein GAN, most GANs fail to obtain the most discriminative projection for real and fake samples in the feature space~\cite{takida2023san}. To address this issue, Takida \textit{et al.}~\cite{takida2023san} proposed a training framework called the slicing adversarial network (SAN) that makes use of features with discriminative projections. They demonstrated its effectiveness in unconditional and class-conditional image generation tasks.

In this paper, we apply the SAN training framework to the vocoding task. It has not yet been verified that SAN is also effective for generating audio waveforms (temporal signals), which have different properties from images (spatial signals). In addition, SAN has not been applied to a case where a neural model is trained with not only an adversarial loss but also auxiliary losses such as regression and feature matching losses. Although most GAN-based vocoders adopt least-squares GAN~\cite{mao2017least}, SAN is not applicable to least-squares GAN in a straightforward way because of the property of its objectives (see Sec.~\ref{sssec:least_square_san}). One possible way to avoid this issue is to replace least-squares GAN with another variant such as minimax GAN~\cite{goodfellow2014generative} or hinge GAN~\cite{lim2017geometric}, but we did not obtain a good result with non-saturating SAN or hinge SAN~\cite{takida2023san} in our preliminary experiments. Indeed, most previous models adopted least-squares GAN because of its training stability~\cite{bollepalli2017generative,pascual17segan,yamamoto2020parallel,kong2020hifigan}, and this is what motivates us to create a SAN counterpart of least-squares GAN. Specifically, we propose a modification scheme called \textit{soft monotonization} to convert least-squares GAN to a SAN model, leading to \textit{least-squares SAN}. Thanks to this technique, \textit{least-squares SAN} can retain the training stability of least-squares GAN. In our experiments, we replace the least-squares GAN utilized in existing vocoders with \textit{least-squares SAN} and demonstrate that SAN can improve the performance of the vocoders, including BigVGAN.
Our contributions are three-fold:
\begin{itemize}
    \setlength{\leftskip}{-15pt}
    \item We propose a \textit{soft monotonization} scheme to modify least-squares GAN to \textit{least-squares SAN}.
    \item We demonstrate that SAN can improve the performance of GAN-based vocoders, including BigVGAN, with small modifications.
    \item We release our code to facilitate reproducibility.
\end{itemize}

\begin{figure*}[htb]
\begin{minipage}[b]{0.3\linewidth}
  \centering
  \centerline{\includegraphics[width=5.5cm]{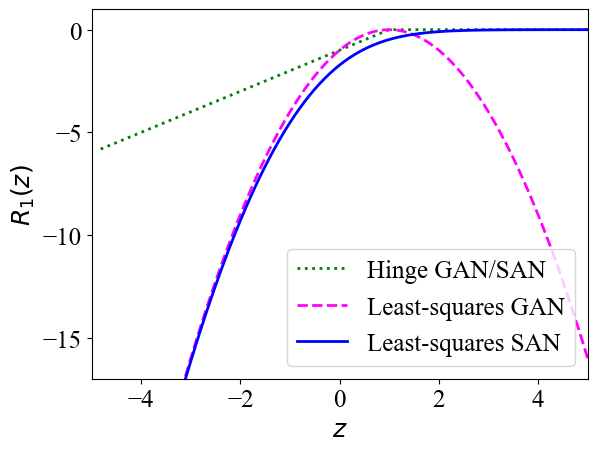}}
  \centerline{(a) $R_1$}\medskip
\end{minipage}
\hfill
\begin{minipage}[b]{.3\linewidth}
  \centering
  \centerline{\includegraphics[width=5.5cm]{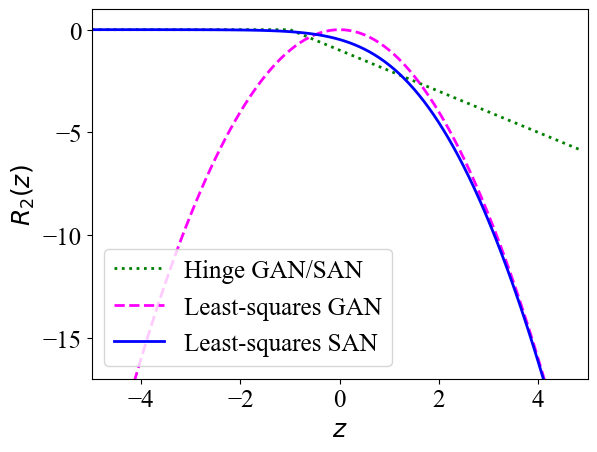}}
  \centerline{(b) $R_2$}\medskip
\end{minipage}
\hfill
\begin{minipage}[b]{0.3\linewidth}
  \centering
  \centerline{\includegraphics[width=5.5cm]{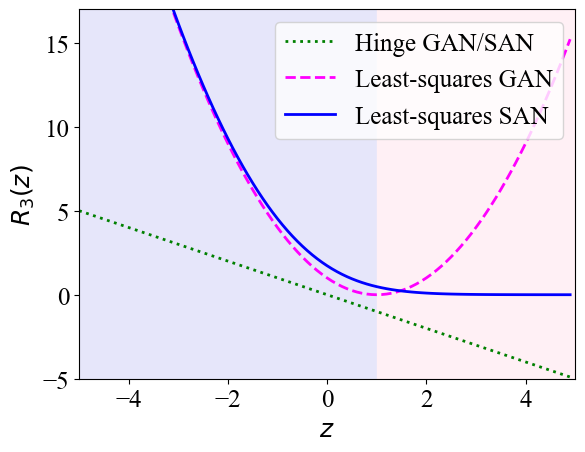}}
  \centerline{(c) $R_3$}\medskip
\end{minipage}
\caption{Comparisons of $R_i:\mathbb{R}\to\mathbb{R}$ $(i=1,2,3)$. In (c), $R_3$ of least-squares GAN is increasing in the red shaded region, which is problematic for SAN due to the non-monotonicity. In contrast, $R_3$ of least-squares SAN is monotonically decreasing over the entire real number but keeps the shape of least-squares GAN to some extent in the blue shaded region.}
\label{fig:comparision_objectives}
\end{figure*}

\section{Related Work}
\label{sec:relateed_work}

GAN was first applied to image generation~\cite{goodfellow2014generative}, and various GANs have achieved impressive results in not only image generation~\cite{karras2019style,sauer2022styleganxl} but also vocoding tasks.

MelGAN~\cite{kumar2019melgan} is a network that introduces a multi-scale discriminator (MSD) to downsample waveforms at multiple scales and applies a window-based discriminator at each scale separately. It also enforces the mapping between a reference waveform and the corresponding generated waveform via a matching loss. Parallel WaveGAN~\cite{yamamoto2020parallel} utilizes a multi-resolution short-term Fourier transform (STFT) loss as an auxiliary loss for GAN training. HiFi-GAN~\cite{kong2020hifigan} introduces a multi-period discriminator (MPD) in addition to the MSD from MelGAN. 
% UnivNet~\cite{jang2021univnet} uses a multi-resolution discriminator (MRD) that takes multi-resolution spectrograms as the input. 
These studies focus primarily on discriminator architectures or auxiliary losses for GAN training. Extending these works, BigVGAN~\cite{lee2023bigvgan} introduces a periodic inductive bias for high-quality waveform synthesis into its generator and scales up the generator, achieving state-of-the-art performance.

In this paper, we apply SAN~\cite{takida2023san}, an improved GAN training framework, to vocoder training. Incorporating the SAN training framework is orthogonal to most types of improvements of discriminator/generator architectures, so we believe our report will contribute to future work in this line of research.

\section{Method}
\label{sec:proposed_method}

\subsection{Overall Framework}
\label{ssec:overall_framework}
The goal of a vocoder is to train a generator function $g_{\bm{\theta}}:S\to X$ that converts mel-spectrogram $\bm{s}\in S$ into waveform signal $\bm{x}\in X$. We focus on GAN-based vocoders and take HiFi-GAN~\cite{kong2020hifigan} and BigVGAN~\cite{lee2023bigvgan} as examples for explanations and discussion. In HiFi-GAN and BigVGAN, multi-scale discriminators, denoted as $\{\bm{\phi}_k\}_{k=1}^K$, are introduced to distinguish ground truth and generated samples. We denote the maximization and minimization objective functions for GAN training as $\mathcal{V}_{\text{GAN}}(\bm{\phi}_k;\bm{\theta})$ and $\Js_{\text{GAN}}(\bm{\theta};\bm{\phi}_k)$, respectively. The intermediate features obtained from the discriminators are also utilized to define pseudo perceptual similarity metrics, a.k.a., feature matching losses $\Js_{\text{FM}}(\bm{\theta};\bm{\phi}_k)$. In addition to these losses, the mel-spectrogram loss $\Js_{\text{mel}}(\bm{\theta})$ is incorporated into the framework to ensure the consistency of given mel-spectrograms and generated waveform signals. To summarize, the overall objective functions can be formulated as
\begin{align}
    &\max_{\{\bm{\phi}_k\}_{k=1}^K}\sum_{k=1}^K\mathcal{V}_{\text{GAN}}(\bm{\phi}_k;\bm{\theta})
    \label{eq:max_problem_vocoder}\\
    &\min_{\bm{\theta}}\sum_{k=1}^K\left(\Js_{\text{GAN}}(\bm{\theta};\bm{\phi}_k)+\lambda_{\text{FM}}\Js_{\text{FM}}(\bm{\theta};\bm{\phi}_k)\right)+\lambda_{\text{mel}}\Js_{\text{mel}}(\bm{\theta}),
    \label{eq:min_problem_vocoder}
\end{align}
where the scalar parameters $\lambda_{\text{FM}}$ and $\lambda_{\text{mel}}$ balance the GAN, feature matching, and mel-spectrogram losses. 
% These parameters need to be carefully tuned.

\subsection{SANs for Vocoder Training}
\label{ssec:san_vocoder}

We derive \textit{least-squares SAN}, a novel variant of adversarial generative models. It enhances discriminators by making their last linear projections more discriminative with only small modifications to least-squares GAN~\cite{mao2017least}.

\begin{table*}[t]
  \centering
  \caption{Objective and subjective evaluations on LibriTTS. Objective results are obtained from a subset of its \texttt{dev} set. Subjective evaluations are based on a 5-scale mean opinion score (MOS) with 95\% confidence interval (CI) from a subset of its \texttt{test} set.}
  \begin{tabular}{c|ccccc|c}
      \bhline{0.8pt}
          Model & M-STFT ($\downarrow$) & PESQ ($\uparrow$) & MCD ($\downarrow$) & Periodicity ($\downarrow$) & V/UV F1 ($\uparrow$) & MOS ($\uparrow$) \\
      \bhline{0.8pt}
          \multicolumn{1}{l|}{Ground truth} & – & – & – & – & – & 3.81$\pm$1.89 \\
      \bhline{0.8pt}
          \multicolumn{1}{l|}{BigVGAN (Lee \textit{et al.}~\cite{lee2023bigvgan})} & 0.7997 & 4.027 & \underline{0.3745} & 0.1018 & 0.9598 & – \\
          \multicolumn{1}{l|}{BigVGAN (our reproduction)} & 0.8382 & 3.862 & \underline{0.3711} & 0.1155 & 0.9540 & 3.19$\pm$2.21 \\
          \multicolumn{1}{l|}{BigVSAN} & \textbf{0.7881} & \underline{4.116} & \textbf{0.3381} & \underline{0.0935} & \underline{0.9635} & \underline{3.24}$\pm$1.95  \\
          \multicolumn{1}{l|}{BigVSAN w/ snakebeta activation} & \underline{0.7992} & \textbf{4.120} & 0.4129 & \textbf{0.0924} & \textbf{0.9644} & \textbf{3.43}$\pm$2.04  \\
      \bhline{0.8pt}
  \end{tabular}
  \label{tb:results_bigvsan}
  \vskip -0.1in
\end{table*}

\subsubsection{From GANs to SANs}
\label{sssec:from_gans_to_sans}

We revisit a sliced optimal transport perspective of GAN optimizations~\cite{takida2023san}, which provides clues for how to enhance GANs.
% A discriminator is usually crafted with a neural network (denoted as $\bm{\phi}$), i.e., a combination of linear operations and non-linear activation functions.
A discriminator function $f:X\to\mathbb{R}$ is usually crafted with a neural network (denoted as $\bm{\phi}$), i.e., a combination of linear operations and non-linear activation functions.
Without loss of generality, we decompose the discriminator into a non-linear function $\bm{h}_{\bm{\varphi}}:X\to W\subseteq\mathbb{R}^D$ and a last linear layer $\bm{w}\in W$ as $f_{\bm{\varphi}}^{\bm{w}}(\bm{x})=\bm{w}^\top \bm{h}_{\bm{\varphi}}(\bm{x})$, i.e., $\bm{\phi}=\{\bm{\varphi},\bm{w}\}$. We can interpret the discrimination process as slicing the non-linear feature $\bm{h}_{\bm{\varphi}}(\bm{x})$ with a common projection $\bm{w}$. Under this parameterization, the maximization and minimization objectives for a discriminator and a generator are respectively formulated as 
\begin{align}
    &\mathcal{V}_{\text{GAN}}(\bm{\phi};\bm{\theta})
    =\E_{p_X}[R_1(f_{\bm{\varphi}}^{\bm{w}}(\bm{x}))]
    +\E_{p_S}[R_2(f_{\bm{\varphi}}^{\bm{w}}(g_{\bm{\theta}}(\bm{s})))]
    \label{eq:max_problem_gan}\\
    &\Js_{\text{GAN}}(\bm{\theta};\bm{\phi})
    =\E_{p_S}[R_3(f_{\bm{\varphi}}^{\bm{w}}(g_{\bm{\theta}}(\bm{s})))],
    \label{eq:min_problem_gan}
\end{align}
where $p_X(\bm{x})$ and $p_S(\bm{s})$ represent the data distributions of waveform signals and mel-spectrograms, respectively.
Regarding the choices of $R_i:\mathbb{R}\to\mathbb{R}$ $(i=1,2,3)$, there are several GAN variants that can be used~\cite{goodfellow2014generative,sebastian2016f,arjovsky2017wasserstein}.
% Through the optimization of the maximization problem, the non-linear function $\bm{h}_{\bm{\varphi}}$ is induced to separate real and fake samples in the feature space $W$, and the projection $\bm{w}$ slices $W$ to discriminate them well.
Through the optimization of the maximization problem, the non-linear function $\bm{h}_{\bm{\varphi}}$ is induced to separate real and fake samples that are mapped onto the feature space $W$ so that they are discriminated well with a linear projection in $W$. However, the linear projection $\bm{w}$ that maximizes Eq.~\beqref{eq:max_problem_gan} cannot make the most of the features for the discrimination~\cite{takida2023san}.

% Takida \textit{et al.}~\cite{takida2023san} found that there are more discriminative linear projections than the optimal $\bm{w}$ of Eq.~\beqref{eq:max_problem_gan} when $\bm{h}_{\bm{\varphi}}$ is given.
% The idea behind SANs is to learn the last projection so that the real and fake samples in $W$ can be accurately distinguished.
% Any GANs can be converted to SANs under the condition where $R_3$, the derivative of which is denoted as $r_3$, is a monotonically decreasing function.
Takida \textit{et al.}~\cite{takida2023san} found that there are more informative linear projections in the sense of the discrimination than the projection $\bm{w}$ that maximizes Eq.~\beqref{eq:max_problem_gan} when $\bm{h}_{\bm{\varphi}}$ is given. They proposed SANs with the aim of learning the last projection layer that can best distinguish real and fake samples in $W$. As long as $R_3$, the derivative of which is denoted as $r_3$, is a monotonically decreasing function, any GANs can be converted to SANs. In other words, the derivative $r_3(z)$ should be negative for any $z\in\mathbb{R}$. Under this condition, the normalization of the last linear layer and the modification of the maximization problem in GANs bring us the objectives for SANs:
\begin{align}
    &\mathcal{V}_{\text{SAN}}(\bm{\varphi},\bm{\omega};\bm{\theta})
    =\E_{p_X}[R_1(f_{\bm{\varphi}}^{\bm{\omega}^-}(\bm{x}))]
    +\E_{p_S}[R_2(f_{\bm{\varphi}}^{\bm{\omega}^-}(g_{\bm{\theta}}(\bm{s})))]\notag\\
    &\qquad\qquad
    \color{purple}{
    -\E_{p_X}[R_3(f_{\bm{\varphi}^-}^{\bm{\omega}}(\bm{x}))]
    +\E_{p_S}[R_3(f_{\bm{\varphi}^-}^{\bm{\omega}}(g_{\bm{\theta}}(\bm{s})))]
    }
    \label{eq:max_problem_san}\\
    &\Js_{\text{SAN}}(\bm{\theta};\bm{\varphi},\bm{\omega})
    =\E_{p_S}[R_3(f_{\bm{\varphi}}^{\bm{\omega}}(g_{\bm{\theta}}(\bm{s})))],
    \label{eq:min_problem_san}
\end{align}
where the last linear layer is normalized onto the unit hypersphere $\mathbb{S}^{D-1}$, i.e., $\bm{\omega}=\bm{w}/\|\bm{w}\|_2$, and $(\cdot)^-$ indicates a parameter with a stop-gradient operator.
Note that the desired direction here depends not on $R_1$ or $R_2$ but on the maximization problem involving $R_3$. Intuitively, the optimal $\bm{\omega}$ best separates $\E_{\tilde{p}_X}[h(\bm{x})]$ and $\E_{\tilde{p}_{\bm{\theta}}}[h(\bm{x})]$, where $\tilde{p}_X(\bm{x})\propto r_3\circ f(\bm{x})p_X(\bm{x})$ and $\tilde{p}_{\bm{\theta}}(\bm{x})\propto r_3\circ f(\bm{x})p_{\bm{\theta}}(\bm{x})$, respectively. Here, $p_{\bm{\theta}}(\bm{x})$ is the generator distribution given by $p_S(\bm{s})$ and the generator function $\bm{x}=g_{\bm{\theta}}(\bm{s})$.

\subsubsection{Soft Monotonization for Least-Squares SAN}
\label{sssec:least_square_san}

Most GAN-based approaches for training vocoders adopt the least-squares GAN objective~\cite{mao2017least}, which is given by
\begin{align}
    R_1(z)=-(1-z)^2\text{, }R_2(z)=-z^2\text{, and } R_3(z)=(1-z)^2.
    \label{eq:rs_lsgan}
\end{align}
As we can see, the function $R_3$ is not a monotonically decreasing function due to the square operator. Therefore, the simple application of the procedure described in Sec.~\ref{sssec:from_gans_to_sans} to least-squares GAN does not lead to a valid SAN model. One possible choice to formulate SAN in our setting is to choose another GAN variant (e.g., hinge GAN or non-saturating GAN). However, our preliminary experiments indicated that the replacement of GAN with hinge SAN or non-saturating SAN~\cite{takida2023san} without extensive hyperparameter tuning led to unsatisfactory results. Indeed, most previous speech-related models adopted least-squares GAN because of its training stability~\cite{bollepalli2017generative,pascual17segan,yamamoto2020parallel,kong2020hifigan}, and this is what motivates us to create a SAN counterpart of least-squares GAN.

Taking into account our motivation, we formulate a variant of SAN that requires only small modifications to least-squares GAN. Specifically, we propose replacing Eq.~\beqref{eq:rs_lsgan} with
\begin{align}
    &\tilde{R}_1(z)=-\varsigma(1-z)^2\text{, }\tilde{R}_2(z)=-\varsigma(z)^2\text{, and}\notag\\
    &\tilde{R}_3(z)=\varsigma(1-z)^2,
    \label{eq:ls_san}
\end{align}
where $\varsigma(\cdot)$ is the softplus function, i.e., $\varsigma(a)=\log(1+e^a)$. We term this replacement \textit{soft monotonization}. Now that $\tilde{R}_3$ satisfies the monotonic condition, we can construct minimax objectives for \textit{least-squares SAN} by substituting Eq.~\beqref{eq:ls_san} into Eqs.~\beqref{eq:max_problem_san} and \beqref{eq:min_problem_san}. Namely, as shown in Fig.~\ref{fig:comparision_objectives}, we design $\tilde{R}_3$ to be a monotonically decreasing function while maintaining the shape of $R_3$ to some extent, which will hopefully retain the least-squares GAN's property regarding training stability. We confirmed the significance of the monotonic property of $\tilde{R}_3$ by observing an unstable training when we adopted Eq.~\beqref{eq:rs_lsgan} instead of Eq.~\beqref{eq:ls_san} in our preliminary experiment. Specifically, loss values tended to get stuck in bad local optima at the early stage of training.

\begin{figure*}[htb]
\begin{minipage}[b]{0.3\linewidth}
  \centering
  \centerline{\includegraphics[width=6.2cm]{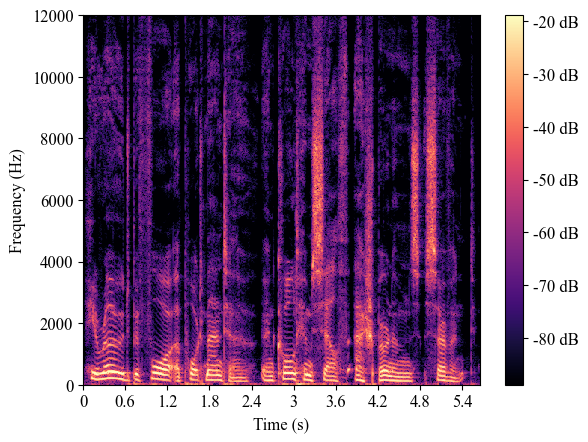}}
  \centerline{(a) Ground truth}\medskip
\end{minipage}
\hfill
\begin{minipage}[b]{.3\linewidth}
  \centering
  \centerline{\includegraphics[width=6.2cm]{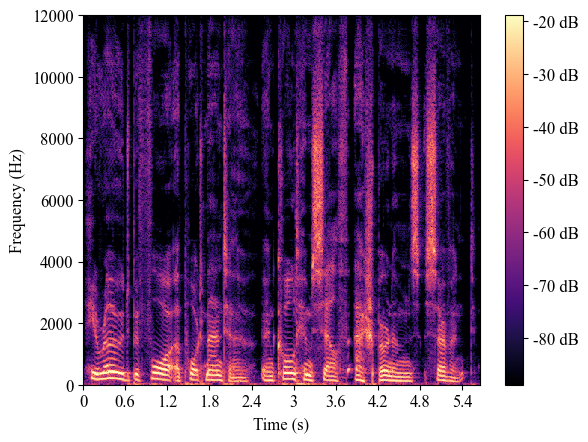}}
  \centerline{(b) BigVSAN}\medskip
\end{minipage}
\hfill
\begin{minipage}[b]{0.3\linewidth}
  \centering
  \centerline{\includegraphics[width=6.2cm]{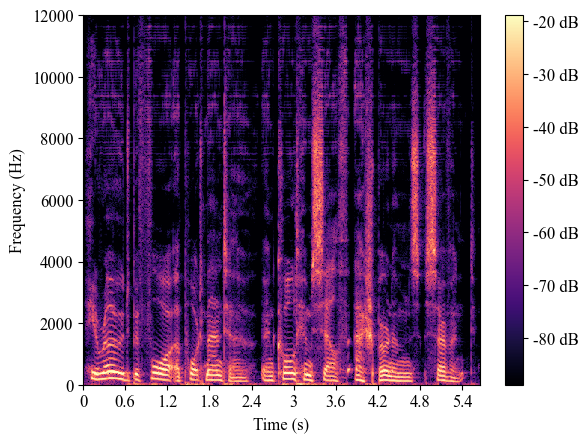}}
  \centerline{(c) BigVSAN w/ snakebeta}\medskip
\end{minipage}
\caption{Spectrograms of synthesized samples with BigVSAN trained on the LibriTTS \texttt{train} set for 1M steps and the corresponding ground truth.}
\label{fig:comparision_spectrograms}
\end{figure*}

\section{Experiments}
\label{sec:experiments}

\subsection{BigVSAN: Large-scale Vocoder Training}

To assess whether SAN is effective in the vocoding task, we apply the SAN training framework (Eqs.~\beqref{eq:max_problem_san} and \beqref{eq:min_problem_san}) to BigVGAN~\cite{lee2023bigvgan}\footnote{\url{https://github.com/NVIDIA/BigVGAN}}, the state-of-the-art large-scale vocoder. We term this vocoder model BigVSAN. We utilize the \texttt{train} set of LibriTTS~\cite{zen2019libritts} (\texttt{train-clean-100}, \texttt{train-clean-360}, and \texttt{train-other-500}) with the original sampling rate of 24 kHz for training, and we follow BigVGAN's preprocessing: 1024 FFT size, 1024 Hann window, 256 hop size, and a 100-band log-mel spectrogram with a frequency range of [0, 12] kHz.

We train three different models as follows. 1) We train a BigVGAN model in our environment to fairly compare BigVGAN and BigVSAN in the same environment. 2) We train a BigVSAN model to assess the effectiveness of the SAN training framework. Since BigVGAN relies on least-squares GAN, we need to introduce \textit{soft monotonization} by applying Eq.~\beqref{eq:ls_san} to train a valid SAN model. 3) We adopt snakebeta activation with the log scale parameterization $f_{\{\alpha,\beta\}}(x)=x+e^{-\beta}\sin^2(e^\alpha x)$, where $\alpha$ and $\beta$ are trainable parameters, which is set as the default in the official repository of BigVGAN. We compare snakebeta activation with snake activation~\cite{ziyin2020neural} $f_\alpha(x)=x+\alpha^{-1}\sin^2(\alpha x)$, where $\alpha$ is trainable, which the above BigVGAN and BigVSAN models utilize. We train all the above models for 1M steps. All other configurations including batch size, optimizer, learning rate scheduler, and balancing parameters of the loss terms follow the official implementation of BigVGAN.

We perform objective evaluations on a subset of the LibriTTS \texttt{dev-clean} and a subset of \texttt{dev-other}. We use 115 audio files selected from \texttt{dev-clean} and 93 files from \texttt{dev-other}, following the official implementation of BigVGAN. We conduct objective evaluations with the following five objective metrics: 1) multi-resolution STFT (M-STFT)~\cite{yamamoto2020parallel}, which measures the spectral distance across multiple resolutions\footnote{\url{https://github.com/csteinmetz1/auraloss}~\cite{steinmetz2020auraloss}}, 2) perceptual evaluation of speech quality (PESQ)~\cite{rix2001pesq}, which is a widely adopted automated assessment of voice quality\footnote{\url{https://github.com/ludlows/PESQ}}, 3) mel-cepstral distortion (MCD)~\cite{kubichek1993mel} with dynamic time warping, which measures the difference between mel cepstra\footnote{\url{https://github.com/ttslr/python-MCD}}, and 4) periodicity error (Periodicity) and 5) F1 score of voiced/unvoiced classification (V/UV F1), which are considered major artifacts from non-autoregressive GAN-based vocoders~\cite{morrison2022chunked}\footnote{\url{https://github.com/descriptinc/cargan}}. All the metrics are first calculated on each of the subsets and then macro-averaged across the subsets. We open-source an all-in-one evaluation tool to facilitate reproducibility at \url{https://github.com/sony/bigvsan_eval}. 
We also perform mean opinion score (MOS) tests on a combined \texttt{test-clean} and \texttt{test-other} set. Eight raters are asked to make quality judgments about synthesized speech samples using five possible responses: 1 = Bad; 2 = Poor; 3 = Fair; 4 = Good; and 5 = Excellent. In total, ten utterances are randomly selected from the combined \texttt{test} set and are then synthesized using the trained models.

Table~\ref{tb:results_bigvsan} lists the results. BigVSAN outperforms BigVGAN in all objective and subjective evaluations when powered by snake activation. SAN successfully boosted the performance of the BigVGAN vocoder, which indicates that SAN can work well in the vocoding task as well as the image generation task. Note that replacing GAN with SAN involves only discriminators and does not introduce any change to generators. Thus, we can keep the same synthesis speed during inference. Next, snakebeta activation deteriorates the MCD and M-STFT scores. This is because snakebeta activation tends to generate artifacts in the high-frequency band, as depicted in Fig.~\ref{fig:comparision_spectrograms}. PESQ considers only the [0, 8] kHz range, but MCD and M-STFT consider both this range and the higher frequency band. Thus, the MCD and M-STFT scores are negatively affected. On the other hand, human raters prefer snakebeta activation to snake activation, and the MOS scores strongly correlate with the PESQ scores, likely because a high-frequency component does not have a large impact on human perception. There is room for discussion on how to evaluate the quality of high-resolution speech samples, especially how largely to consider artifacts in the high-frequency band. We leave it for future work. Besides, the 95\% confidence intervals in the MOS assessment are quite large. We hypothesize this is because differences in generation quality between different vocoders (including recorded ground truth) are small. The raters might have struggled to find differences between provided samples and then sometimes gave low scores to ground-truth samples. This matches the experimental results in a previous study~\cite{cooper23investigating}. Audio samples are available at \url{https://TakashiShibuyaSony.github.io/bigvsan/}.

\begin{table}[t]
  \centering
  \caption{Results for Fr\'{e}chet Audio Distance (FAD) evaluated on three datasets: LJ Speech, LibriTTS, and VCTK. Scores marked with $\dagger$ are reported in the VocBench paper~\cite{albadawy2022vocbench}.}
  \resizebox{\columnwidth}{!}{
  \begin{tabular}{c|cc|cc}
      \bhline{0.8pt}
          \multirow{2}{*}{\textbf{Dataset}} & \multirow{2}{*}{\textbf{MelGAN}$^\dagger$} & \multirow{2}{*}{\textbf{MelSAN}} & \textbf{Parallel} & \textbf{Parallel} \\
           & & & \textbf{WaveGAN}$^\dagger$ & \textbf{WaveSAN} \\
      \bhline{0.8pt}
          LJ Speech & 1.51 & \textbf{1.34} & 0.92 & \textbf{0.84} \\
          LibriTTS & 2.95 & \textbf{2.91} & 1.41 & \textbf{0.87} \\
          VCTK & 1.76 & \textbf{1.69} & 1.22 & \textbf{0.76} \\
      \bhline{0.8pt}
  \end{tabular}
  }
  \label{tb:results_vocbench}
  \vskip -0.1in
\end{table}

\subsection{Moderate-sized Vocoder Training}

To further investigate the effectiveness of the SAN training framework, we apply SAN to moderate-sized neural vocoders. For this purpose, we utilize the VocBench framework~\cite{albadawy2022vocbench}\footnote{\url{https://github.com/facebookresearch/vocoder-benchmark}}. VocBench provides a shared environment where we can train and evaluate different neural vocoders on three public datasets: LJ Speech~\cite{ito2017ljspeech}, LibriTTS~\cite{zen2019libritts}, and VCTK~\cite{yamagishi2019vctk}. 
% LJ Speech consists of 13,100 audio clips of a single speaker reading passages from non-fiction books. In VocBench, the first 20 clips are used as a test split, the following 10 clips as a validation split, and the rest as a training split. Regarding LibriTTS, the \texttt{train-clean-100} and \texttt{train-clean-360} subsets are used for training, \texttt{dev-clean} for validation, and \texttt{test-clean} for testing in VocBench. VCTK is a multi-speaker dataset. Each speaker reads out about 400 sentences selected from a newspaper, the rainbow passage, and an elicitation paragraph utilized for the speech accent archive. In VocBench, a randomly selected 85\% of the samples are used as a training set, 10\% as a validation set, and 5\% for a test set. See the original paper~\cite{albadawy2022vocbench} for more details.
See the original paper~\cite{albadawy2022vocbench} for details, including training/validation/test splits.

In this experiment, we apply SAN to two different GAN-based neural vocoders: MelGAN~\cite{kumar2019melgan} and Parallel WaveGAN~\cite{yamamoto2020parallel}. We call the SAN-applied models MelSAN and Parallel WaveSAN, respectively. Since both vocoders implemented in the VocBench framework rely on least-squares GAN, we apply \textit{soft monotonization}. We train the MelSAN and Parallel WaveSAN models with the same hyperparameter settings as the GAN counterparts implemented in VocBench. In other words, we only replace least-squares GAN with \textit{least-squares SAN} on each vocoder. We evaluate trained neural vocoders using Fréchet Audio Distance (FAD)~\cite{kilgour2019frechet}, as FAD scores strongly correlate with subjective evaluations~\cite{albadawy2022vocbench}.

Table~\ref{tb:results_vocbench} lists the results. As we can see, SAN outperforms GAN in all combinations of vocoder model and dataset. This demonstrates that SAN can improve the performance of a neural vocoder in various situations with only simple modifications to GAN.

\section{Conclusion}
\label{sec:conclusion}

In this paper, we investigated the effectiveness of the SAN training framework in the vocoding task. We proposed \textit{soft monotonization} to convert least-squares GAN to a SAN model. 
% Our experimental results demonstrated that SAN can boost even the performance of BigVGAN, the state-of-the-art vocoder, with just a few modifications. In addition, we empirically verified that SAN can improve the performance of various vocoders with various datasets. 
Our experimental results demonstrated that SAN can improve the performance of various vocoders, including BigVGAN, with various datasets. 
We obtained the results without extra hyperparameter tuning thanks to our proposed \textit{least-squares SAN}. Since incorporating the SAN training framework is orthogonal to most types of improvements of discriminator/generator architectures, we believe that SAN can boost other GAN-based vocoders and further contribute to this line of research.

\vspace{3mm}\noindent\textbf{Acknowledgement.} Computational resource of AI Bridging Cloud Infrastructure (ABCI) provided by National Institute of Advanced Industrial Science and Technology (AIST) was used.

\clearpage

% References should be produced using the bibtex program from suitable
% BiBTeX files (here: strings, refs, manuals). The IEEEbib.bst bibliography
% style file from IEEE produces unsorted bibliography list.
% -------------------------------------------------------------------------
\vfill\pagebreak
\bibliographystyle{IEEEbib}
\bibliography{str_def_abrv,refs_dgm,refs_ml,refs_speech}

\end{document}